\documentclass[aps,prb,twocolumn,showpacs]{revtex4}

\usepackage{graphicx}
\usepackage{amsmath}
\usepackage{amssymb}

\begin{document}

\title{Out-of-plane impurities induced the deviation from the
monotonic d-wave superconducting gap in cuprate superconductors}


\author{Zhi Wang and Shiping Feng$^{*}$}
\affiliation{Department of Physics, Beijing Normal University,
Beijing 100875, China}

\begin{abstract}
The electronic structure of cuprate superconductors is studied
within the kinetic energy driven superconducting mechanism in the
presence of out-of-plane impurities. With increasing impurity
concentration, although both superconducting coherence peaks around
the nodal and antinodal regions are suppressed, the position of the
leading-edge mid-point of the electron spectrum around the nodal
region remains at the same position, whereas around the antinodal
region it is shifted towards higher binding energies, this leads to
a strong deviation from the monotonic d-wave superconducting gap in
the out-of-plane impurity-controlled cuprate superconductors.
\end{abstract}


\maketitle

\section{Introduction}

The superconducting gap is a fundamental property of superconductors
\cite{schrieffer}, and the nature of its anisotropy has played a
crucial role in the testing of the microscopic theory of
superconductivity in cuprate superconductors \cite{anderson}.
Experimentally, by virtue of systematic measurements \cite{tsuei},
particularly using the angle-resolved photoemission spectroscopy
(ARPES) technique \cite{shen}, the d-wave nature of the
superconducting gap has been well established by now. In particular,
this d-wave superconducting symmetry remains one of the cornerstones
of our understanding of the physics in cuprate superconductors
\cite{shen,tsuei,perali,sangiovanni,zhang}. The early ARPES
measurements on the cuprate superconductor
Bi$_{2}$Sr$_{2}$CaCu$_{2}$O$_{8+\delta}$ \cite{shi} showed that in
the real space the gap function and the pairing force have a range
of one lattice spacing, and then the superconducting gap function is
of the monotonic d-wave form $\Delta_{\bf k}= \Delta [{\rm cos}k_{x}
-{\rm cos} k_{y}]/2$. Later, the ARPES measurements on the cuprate
superconductor Bi$_{2}$Sr$_{2}$CaCu$_{2}$O$_{8+\delta}$ \cite{mesot}
indicated that the superconducting gap significantly deviates from
this monotonic d-wave form. Furthermore, it was argued that this
deviation should be attributed to an increase of the electron
correlation, which may increase the intensity of the higher order of
the harmonic component in the monotonic d-wave gap function
\cite{mesot}. However, recent ARPES measurements
\cite{kondo,hashimoto} on the cuprate superconductors
(Bi,Pb)$_{2}$(Sr,La)$_{2}$CuO$_{6+\delta}$ and
Bi$_{2}$Sr$_{1.6}$$Ln$$_{0.4}$CuO$_{6+\delta}$ ($Ln$-Bi2201) with
$Ln$=La, Nd, and Gd showed that a much stronger deviation from the
monotonic d-wave superconducting gap form is unlikely to be due to
the strong correlation effect.

The cuprate superconductors have a layered structure consisting of
the two-dimensional CuO$_{2}$ layers separated by insulating layers
\cite{bednorz,kastner}. The single common feature is the presence of
the CuO$_{2}$ plane \cite{shen,kastner}, and it seems evident that
the unusual behaviors of cuprate superconductors are dominated by
this CuO$_{2}$ plane \cite{anderson}. It has been well established
that the Cu$^{2+}$ ions exhibit an antiferromagnetic long-range
order in the parent compounds of cuprate superconductors, and
superconductivity occurs when the antiferromagnetic long-range order
state is suppressed by doped charge carriers \cite{kastner}. Since
these doped charge carriers are induced by the replacement of ions
by those with different valences or the addition of excess oxygens
in the block layer, therefore in principle, all cuprate
superconductors have naturally impurities (or disorder). However,
for the cuprate superconductors
(Bi,Pb)$_{2}$(Sr,La)$_{2}$CuO$_{6+\delta}$ and $Ln$-Bi2201, the
mismatch in the ionic radius between Bi and Pb or Sr and $Ln$ causes
the {\it out-of-plane impurities} \cite{eisaki}, where the
concentration of the out-of-plane impurities is controlled by
varying the radius of the Pb or $Ln$ ions, and then the
superconducting transition temperature $T_{c}$ is found to be
decreasing with increasing impurity concentration. These cuprate
superconductors (Bi,Pb)$_{2}$(Sr,La)$_{2}$CuO$_{6+\delta}$ and
$Ln$-Bi2201 are often referred to as the out-of-plane
impurity-controlled cuprate superconductors. Recently, the
electronic structure of the out-of-plane impurity-controlled cuprate
superconductors and the related superconducting gap function have
been investigated experimentally by using ARPES
\cite{kondo,hashimoto}. It was shown that although the effect of the
out-of-plane impurity scattering around the antinodal region is much
stronger than that around the nodal region, both superconducting
coherence peaks around the nodal and antinodal regions are
suppressed. Furthermore, the magnitude of the deviation from the
monotonic d-wave superconducting gap form increases with increasing
impurity concentration \cite{kondo,hashimoto}. The appearance of the
strong deviation from the monotonic d-wave superconducting gap form
observed recently in the out-of-plane impurity-controlled cuprate
superconductors (Bi,Pb)$_{2}$(Sr,La)$_{2}$CuO$_{6+\delta}$ and
$Ln$-Bi2201 is the most remarkable effect \cite{kondo,hashimoto},
however, its full understanding is still a challenging issue. To the
best of our knowledge, this strong deviation from the monotonic
d-wave superconducting gap form in the out-of-plane
impurity-controlled cuprate superconductors has not been treated
starting from a microscopic superconducting theory yet.

In the absence of out-of-plane impurity scattering, the electronic
structure of cuprate superconductors in the superconducting state
has been discussed \cite{guo,feng} within the framework of the
kinetic energy driven superconductivity \cite{feng1}, where the
superconducting gap function has a monotonic d-wave form, and the
main features of the ARPES experiments \cite{shen} on cuprate
superconductors have been reproduced. In this paper, we study the
electronic structure of the out-of-plane impurity-controlled cuprate
superconductors in the superconducting state and the related
superconducting gap function along with this line. We employ the
$t$-$J$ model by considering the out-of-plane impurity scattering,
and then show explicitly that the strong deviation from the
monotonic d-wave superconducting gap form occurs due to the presence
of the impurity scattering. Although both sharp superconducting
coherence peaks around the nodal and the antinodal regions are
suppressed, the effect of the impurity scattering is stronger in the
antinodal region than that in the nodal region. Our results also
show that the electronic structure of the out-of-plane
impurity-controlled cuprate superconductors in the superconducting
state can be understood within the framework of the kinetic energy
driven superconducting mechanism with the out-of-plane impurity
scattering taken into account.

This paper is organized as follows. In Sec. II we present the basic
formalism of the electronic structure calculation in the presence of
the out-of-plane impurities. Within this theoretical framework, we
discuss the electronic structure of the out-of-plane
impurity-controlled cuprate superconductors in the superconducting
state and the related superconducting gap function in Sec. III,
where we show that the well pronounced deviation from the monotonic
d-wave superconducting gap form is mainly caused by the out-of-plane
impurity scattering. Finally, we give a summary in Sec. IV.

\section{Formalism}

It has been shown that the essential physics of cuprate
superconductors is properly accounted by the two-dimensional $t$-$J$
model on a square lattice \cite{anderson},
\begin{eqnarray}
H&=&-t\sum_{i\hat{\eta}\sigma}C^{\dagger}_{i\sigma}
C_{i+\hat{\eta}\sigma}+t'\sum_{i\hat{\tau}\sigma}
C^{\dagger}_{i\sigma}C_{i+\hat{\tau}\sigma}+\mu\sum_{i\sigma}
C^{\dagger}_{i\sigma}C_{i\sigma}\nonumber\\
&+&J\sum_{i\hat{\eta}}{\bf S}_{i} \cdot {\bf S}_{i+\hat{\eta}},
\end{eqnarray}
acting on the Hilbert subspace with no doubly occupied site, i.e.,
$\sum_{\sigma}C^{\dagger}_{i\sigma}C_{i\sigma}\leq 1$, where
$\hat{\eta}=\pm\hat{x},\pm \hat{y}$, $\hat{\tau}=\pm\hat{x}
\pm\hat{y}$, $C^{\dagger}_{i\sigma}$ ($C_{i\sigma}$) is the creation
(annihilation) operator of an electron with spin $\sigma$, ${\bf
S}_{i}=(S^{x}_{i},S^{y}_{i}, S^{z}_{i})$ are spin operators, and
$\mu$ is the chemical potential. To deal with the constraint of no
double occupancy in analytical calculations, the charge-spin
separation fermion-spin theory \cite{feng2} has been developed,
where the constrained electron operators $C_{i\uparrow}$ and
$C_{i\downarrow}$ are decoupled as
$C_{i\uparrow}=h^{\dagger}_{i\uparrow}S^{-}_{i}$ and
$C_{i\downarrow}=h^{\dagger}_{i\downarrow}S^{+}_{i}$, respectively,
here the spinful fermion operator
$h_{i\sigma}=e^{-i\Phi_{i\sigma}}h_{i}$ describes the charge degree
of freedom together with some effects of spin configuration
rearrangements due to the presence of the doped charge carrier
itself, while the spin operator $S_{i}$ describes the spin degree of
freedom, then the electron on-site local constraint for the single
occupancy, $\sum_{\sigma}C^{\dagger}_{i\sigma}
C_{i\sigma}=S^{+}_{i}h_{i\uparrow}h^{\dagger}_{i\uparrow}S^{-}_{i}
+S^{-}_{i}h_{i\downarrow}h^{\dagger}_{i\downarrow}S^{+}_{i}=h_{i}
h^{\dagger}_{i}(S^{+}_{i}S^{-}_{i}+S^{-}_{i}S^{+}_{i})=1-
h^{\dagger}_{i}h_{i}\leq 1$, is satisfied in analytical
calculations. In particular, it has been shown that under the
decoupling scheme, this charge-spin separation fermion-spin
representation is a natural representation of the constrained
electron defined in the Hilbert subspace without double electron
occupancy \cite{feng}. Furthermore, these charge carrier and spin
are gauge invariant \cite{feng2}, and in this sense they are real
and can be interpreted as physical excitations \cite{laughlin}. In
this charge-spin separation fermion-spin representation, the $t$-$J$
model (1) can be expressed as,
\begin{eqnarray}
H&=&t\sum_{i\hat{\eta}}(h^{\dagger}_{i+\hat{\eta}\uparrow}
h_{i\uparrow}S^{+}_{i}S^{-}_{i+\hat{\eta}}+
h^{\dagger}_{i+\hat{\eta} \downarrow}h_{i\downarrow}S^{-}_{i}
S^{+}_{i+\hat{\eta}})\nonumber\\
&-&t'\sum_{i\hat{\tau}}(h^{\dagger}_{i+\hat{\tau}\uparrow}
h_{i\uparrow}S^{+}_{i}S^{-}_{i+\hat{\tau}}+
h^{\dagger}_{i+\hat{\tau}\downarrow}h_{i\downarrow}S^{-}_{i}
S^{+}_{i+\hat{\tau}})\nonumber \\
&-&\mu\sum_{i\sigma}h^{\dagger}_{i\sigma} h_{i\sigma}+J_{{\rm eff}}
\sum_{i\hat{\eta}}{\bf S}_{i}\cdot {\bf S}_{i+\hat{\eta}},
\end{eqnarray}
with $J_{{\rm eff}}=(1-\delta)^{2}J$, and $\delta=\langle
h^{\dagger}_{i\sigma}h_{i\sigma}\rangle=\langle h^{\dagger}_{i}
h_{i}\rangle$ being the charge carrier doping concentration. This
$J_{{\rm eff}}$ is similar to that obtained in Gutzwiller approach
\cite{zhang}. As an important consequence, the kinetic energy term
in the $t$-$J$ model has been transferred as the interaction between
charge carriers and spins, which reflects that even the kinetic
energy term in the $t$-$J$ Hamiltonian has a strong Coulombic
contribution due to the restriction of no double occupancy of a
given site. This interaction from the kinetic energy term is quite
strong, and it has been shown \cite{feng1} in terms of the
Eliashberg's strong coupling theory \cite{eliashberg} that in the
case without an antiferromagnetic long-range order, this interaction
can induce a charge carrier pairing state (then the electron Cooper
pairing state) with d-wave symmetry by exchanging spin excitations
in the higher power of the charge carrier doping concentration
$\delta$. In this case, the electron Cooper pairs originating from
the charge carrier pairing state are due to the charge-spin
recombination, and their condensation reveals the d-wave
superconducting ground-state. Furthermore, this d-wave
superconducting state is controlled by both the superconducting gap
function and the quasiparticle coherence, which leads to the fact
that the maximal superconducting transition temperature occurs
around the optimal doping, and then decreases in both underdoped and
overdoped regimes \cite{feng1}. Moreover, it has been shown
\cite{guo,feng} that this superconducting state is the conventional
Bardeen-Cooper-Schrieffer (BCS) like \cite{schrieffer,bcs} with the
d-wave symmetry, so that the basic BCS formalism with the d-wave
superconducting gap function is still valid in quantitatively
reproducing all main low energy features of the ARPES experimental
measurements on cuprate superconductors, although the pairing
mechanism is driven by the kinetic energy by exchanging spin
excitations, and other exotic magnetic scattering \cite{dai} is
beyond the BCS formalism. Following previous discussions
\cite{guo,feng,feng1}, the full charge carrier Green's function in
the superconducting state with a monotonic d-wave gap function can
be obtained in the Nambu representation as \cite{wang},
\begin{eqnarray}
\tilde{g}({\bf k},\omega)&=&Z_{hF}{1\over\omega^{2}-E^{2}_{h{\bf
k}}} \left(
\begin{array}{cc}
{\omega+\bar{\xi}_{{\bf k}}} & {\bar{\Delta}_{hZ}({\bf k})} \\
{\bar{\Delta}_{hZ}({\bf k})} & {\omega-\bar{\xi}_{{\bf k}}}
\end{array}\right)\nonumber\\
&=&Z_{hF}{\omega\tau_{0}+\bar{\Delta}_{hZ} ({\bf k})
\tau_{1}+\bar{\xi}_{{\bf k}}\tau_{3}\over\omega^{2}- E^{2}_{h{\bf k}
}},
\end{eqnarray}
where $\tau_{0}$ is the unit matrix, $\tau_{1}$ and $\tau_{3}$ are
the Pauli matrices, the renormalized charge carrier excitation
spectrum $\bar{\xi}_{{\bf k}}=Z_{hF}\xi_{\bf k}$, with the
mean-field charge carrier excitation spectrum $\xi_{{\bf k}}
=Zt\chi_{1} \gamma_{{\bf k}}-Zt'\chi_{2}\gamma'_{{\bf k}}-\mu$, the
spin correlation functions $\chi_{1}=\langle
S_{i}^{+}S_{i+\hat{\eta} }^{-}\rangle$ and $\chi_{2}=\langle
S_{i}^{+}S_{i+\hat{\tau}}^{-} \rangle$, $\gamma_{{\bf
k}}=(1/Z)\sum_{\hat{\eta}}e^{i{\bf k}\cdot \hat{\eta}}$,
$\gamma'_{{\bf k}}=(1/Z)\sum_{\hat{\tau}}e^{i{\bf k}
\cdot\hat{\tau}}$, $Z$ is the number of the nearest neighbor or next
nearest neighbor sites, the renormalized charge carrier monotonic
d-wave pair gap function $\bar{\Delta}_{hZ}({\bf k})=Z_{hF}
\bar{\Delta}_{h}({\bf k})$, where the effective charge carrier
monotonic d-wave pair gap function $\bar{\Delta}_{h}({\bf k})=
\bar{\Delta}_{h}\gamma^{(d)}_{{\bf k}}$ with $\gamma^{(d)}_{{\bf k}}
=({\rm cos} k_{x}-{\rm cos}k_{y})/2$, and the charge carrier
quasiparticle spectrum $E_{h{\bf k}}=\sqrt {\bar{\xi}^{2}_{{\bf k}}
+\mid\bar{\Delta}_{hZ}({\bf k})\mid^{2}}$. The charge carrier
quasiparticle coherent weight $Z_{hF}$ and effective charge carrier
gap parameter $\bar{\Delta}_{h}$ are determined by the following two
equations \cite{guo,feng,feng1},
\begin{widetext}
\begin{subequations}
\begin{eqnarray}
1&=&{1\over N^{3}}\sum_{{\bf k,p,p'}}\Lambda^{2}_{{\bf p+k}}
\gamma^{(d)}_{{\bf k-p'+p}}\gamma^{(d)}_{{\bf k}}{Z^{2}_{hF}\over
E_{h{\bf k}}}{B_{{\bf p}}B_{{\bf p'}}\over\omega_{{\bf p}}
\omega_{{\bf p'}}}\left({F^{(1)}_{1}({\bf k,p,p'})\over
(\omega_{{\bf p'}} -\omega_{{\bf p}})^{2}-E^{2}_{h{\bf k}}}-{
F^{(2)}_{1}({\bf k,p,p'}) \over(\omega_{{\bf p'}}+
\omega_{{\bf p}})^{2}-E^{2}_{h{\bf k}}}\right ),\\
{1\over Z}_{hF} &=& 1+{1\over N^{2}}\sum_{{\bf p,p'}}
\Lambda^{2}_{{\bf p}+{\bf k}_{0}}Z_{hF}{B_{{\bf p}}B_{{\bf p'}}\over
4\omega_{{\bf p}}\omega_{{\bf p'}}}\left({F^{(1)}_{2}({\bf p,p'})
\over(\omega_{{\bf p}}- \omega_{{\bf p'}}-E_{h{\bf p-p'+k_{0}}}
)^{2}}+{F^{(2)}_{2}({\bf p,p'})\over(\omega_{{\bf p}}-\omega_{{\bf
p'}}+E_{h{\bf p-p'+k_{0}}})^{2}}\right . \nonumber \\
&+& \left . {F^{(3)}_{2}({\bf p,p'})\over (\omega_{{\bf p}}+
\omega_{{\bf p'}}-E_{h{\bf p-p'+k_{0}}})^{2}}+{F^{(4)}_{2} ({\bf p,
p'})\over(\omega_{{\bf p}}+\omega_{{\bf p'}}+E_{h{\bf p-p'+k_{0}}}
)^{2}}\right ) ,
\end{eqnarray}
\end{subequations}
\end{widetext}
respectively, where ${\bf k}_{0}=[\pi,0]$, $\Lambda_{\bf k}=Zt
\gamma_{\bf k}-Zt'\gamma'_{\bf k}$, $B_{{\bf p}}=2\lambda_{1}(A_{1}
\gamma_{{\bf p}}-A_{2})-\lambda_{2}(2\chi^{z}_{2}\gamma_{{\bf p }}'-
\chi_{2})$, $\lambda_{1}=2ZJ_{{\rm eff}}$, $\lambda_{2}=4Z\phi_{2}
t'$, $A_{1}=\epsilon \chi^{z}_{1}+\chi_{1}/2$, $A_{2}=\chi^{z}_{1}+
\epsilon\chi_{1}/2$, $\epsilon=1+2t\phi_{1}/J_{{\rm eff}}$, the
charge carrier's particle-hole parameters $\phi_{1}=\langle
h^{\dagger}_{i\sigma}h_{i+\hat{\eta}\sigma}\rangle$ and $\phi_{2}=
\langle h^{\dagger}_{i\sigma} h_{i+\hat{\tau}\sigma}\rangle$, the
spin correlation functions $\chi^{z}_{1}=\langle S_{i}^{z}
S_{i+\hat{\eta}}^{z}\rangle$ and $\chi^{z}_{2}=\langle S_{i}^{z}
S_{i+\hat{\tau}}^{z}\rangle$, $F^{(1)}_{1}({\bf k,p,p'})=
(\omega_{{\bf p'}}-\omega_{{\bf p}})[n_{B}(\omega_{{\bf p}})-
n_{B}(\omega_{{\bf p'}})][1-2n_{F}(E_{h{\bf k}})]+E_{h{\bf k}}[n_{B}
(\omega_{{\bf p'}})n_{B}(-\omega_{{\bf p}})+n_{B}(\omega_{{\bf p}})
n_{B}(-\omega_{{\bf p'}})]$, $F^{(2)}_{1}({\bf k, p,p'})=
(\omega_{{\bf p'}}+\omega_{{\bf p}})[n_{B}(-\omega_{{\bf p'}})-
n_{B}(\omega_{{\bf p}})][1-2n_{F}(E_{h{\bf k}})]+E_{h{\bf k}}[n_{B}
(\omega_{{\bf p'}})n_{B}(\omega_{{\bf p}})+n_{B}(-\omega_{{\bf p'}})
n_{B}(-\omega_{{\bf p}})]$, $F^{(1)}_{2}({\bf p,p'})=n_{F}(E_{h{\bf
p-p'+k_{0}}})[n_{B}(\omega_{{\bf p'}})-n_{B} (\omega_{{\bf p}})]-
n_{B}(\omega_{{\bf p}})n_{B}(-\omega_{{\bf p'}} )$, $F^{(2)}_{2}
({\bf p,p'})=n_{F}(E_{h{\bf p-p'+k_{0}}}) [n_{B} (\omega_{{\bf p}})
-n_{B}(\omega_{{\bf p'}})]-n_{B}(\omega_{{\bf p'}} )n_{B}
(-\omega_{{\bf p}})$, $F^{(3)}_{2}({\bf p,p'})=n_{F}(E_{h{\bf p-p'+
k_{0}}})[n_{B}(\omega_{{\bf p'}})-n_{B}(-\omega_{{\bf p}})]+ n_{B}
(\omega_{{\bf p}})n_{B}(\omega_{{\bf p'}})$, $F^{(4)}_{2}({\bf p,p'}
)=n_{F}(E_{h{\bf p-p'+k_{0}}})[n_{B} (-\omega_{{\bf p'}})-n_{B}
(\omega_{{\bf p}})]+n_{B}(-\omega_{{\bf p}})n_{B}(-\omega_{{\bf p'}}
)$, $n_{B}(\omega_{{\bf p}})$ and $n_{F}(E_{h{\bf k}})$ are the
boson and fermion distribution functions, respectively, and the
mean-field spin excitation spectrum,
\begin{widetext}
\begin{eqnarray}
\omega^{2}_{{\bf p}}&=&\lambda_{1}^{2}\left[\left( A_{4}-\alpha
\epsilon\chi^{z}_{1}\gamma_{{\bf p}}-{1\over 2Z}\alpha\epsilon
\chi_{1}\right)(1-\epsilon\gamma_{{\bf p}})+{1\over 2}\epsilon
\left(A_{3}-{1\over 2} \alpha \chi^{z}_{1}-\alpha \chi_{1}
\gamma_{{\bf p}}\right)(\epsilon-\gamma_{{\bf p}})\right]
+\lambda_{2}^{2}\left[\alpha\left(\chi^{z}_{2}\gamma_{{\bf p}}'-
{3\over 2Z}\chi_{2}\right)\gamma_{{\bf p}}'\right. \nonumber\\
&+&\left. {1\over 2}\left(A_{5}- {1\over 2}\alpha\chi^{z}_{2}
\right)\right]+\lambda_{1}\lambda_{2} \left[ \alpha\chi^{z}_{1}
(1-\epsilon \gamma_{{\bf p}})\gamma_{{\bf p}}' +{1\over 2}\alpha
(\chi_{1} \gamma_{{\bf p}}'-C_{3})(\epsilon-\gamma_{{\bf p}})+
\alpha \gamma_{{\bf p}}'(C^{z}_{3} -\epsilon \chi^{z}_{2}
\gamma_{{\bf p}})-{1\over 2}\alpha \epsilon(C_{3}- \chi_{2}
\gamma_{{\bf p}})\right],~~~~~
\end{eqnarray}
\end{widetext}
where $A_{3}=\alpha C_{1}+(1-\alpha)/(2Z)$, $A_{4}=\alpha C^{z}_{1}
+(1-\alpha)/(4Z)$, $A_{5}=\alpha C_{2}+(1-\alpha)/(2Z)$, and the
spin correlation functions
$C_{1}=(1/Z^{2})\sum_{\hat{\eta},\hat{\eta'}}\langle
S_{i+\hat{\eta}}^{+}S_{i+\hat{\eta'}}^{-}\rangle$,
$C^{z}_{1}=(1/Z^{2})\sum_{\hat{\eta},\hat{\eta'}}\langle
S_{i+\hat{\eta}}^{z}S_{i+\hat{\eta'}}^{z}\rangle$,
$C_{2}=(1/Z^{2})\sum_{\hat{\tau},\hat{\tau'}}\langle
S_{i+\hat{\tau}}^{+}S_{i+\hat{\tau'}}^{-}\rangle$,
$C_{3}=(1/Z)\sum_{\hat{\tau}}\langle S_{i+\hat{\eta}}^{+}
S_{i+\hat{\tau}}^{-}\rangle$, and $C^{z}_{3}=(1/Z)
\sum_{\hat{\tau}}\langle S_{i+\hat{\eta}}^{z}
S_{i+\hat{\tau}}^{z}\rangle$. In order to satisfy the sum rule of
the correlation function $\langle S^{+}_{i}S^{-}_{i}\rangle=1/2$ in
the case without the antiferromagnetic long-range order, an
important decoupling parameter $\alpha$ has been introduced in the
above calculation \cite{guo,feng,feng1}, which can be regarded as
the vertex correction. These two equations (4a) and (4b) must be
solved simultaneously with other self-consistent equations, then all
order parameters, the decoupling parameter $\alpha$, and the
chemical potential $\mu$ are determined by the self-consistent
calculation \cite{guo,feng,feng1}. In this sense, the calculations
in this kinetic energy driven superconductivity scheme are
controllable without using any adjustable parameters. We emphasize
that the Green's function (3) is obtained within the kinetic energy
driven superconducting mechanism, although the similar
phenomenological expression has been used to discuss the impurity
effect in cuprate superconductors \cite{haas,graser}.

With the charge carrier BCS formalism (3) under the kinetic energy
driven superconducting mechanism, we can now introduce the effect of
impurity scatterers into the electronic structure. In the presence
of impurities, the unperturbed charge carrier Green's function in
Eq. (3) is dressed by impurity scattering \cite{wang},
\begin{eqnarray}
\tilde{g}_{I}({\bf k},\omega)^{-1}=\tilde{g}({\bf k},\omega)^{-1}-
Z_{hF}^{-1}\tilde{\Sigma}({\bf k},\omega),
\end{eqnarray}
with the self-energy function $\tilde{\Sigma}({\bf k},\omega)=
\sum_{\alpha}\Sigma_{\alpha}({\bf k},\omega)\tau_{\alpha}$. In this
case, the charge carrier Green's function in Eq. (6) can be
explicitly rewritten as,
\begin{widetext}
\begin{eqnarray}
\tilde{g}_{I}({\bf k},\omega)=\sum_{\alpha}g_{I\alpha}({\bf k}
,\omega)\tau_{\alpha} =Z_{hF}{[\omega-\Sigma_{0}({\bf k},\omega)]
\tau_{0}+ [\bar{\Delta}_{hZ}({\bf k})+\Sigma_{1}({\bf
k},\omega)]\tau_{1}+ [\bar{\xi}_{{\bf k}}+\Sigma_{3}({\bf
k},\omega)]\tau_{3} \over [\omega-\Sigma_{0}({\bf k},\omega)]^{2}
-[\bar{\xi}_{{\bf k}}+ \Sigma_{3}({\bf
k},\omega)]^2-[\bar{\Delta}_{hZ}({\bf k})+ \Sigma_{1}({\bf
k},\omega)]^2}.
\end{eqnarray}
\end{widetext}
Based on this Green's function (7), we \cite{wang} have discussed
the effect of the extended impurity scatterers on the quasiparticle
transport of cuprate superconductors in the superconducting state
within the nodal approximation of the quasiparticle excitations and
scattering processes, where the main effect on the quasiparticle
transport comes from the extended impurity {\it forward} (or {\it
diagonal}) {\it scatterers}, and therefore the component of the
self-energy function $\Sigma_{1}({\bf k},\omega)$ has been
neglected, while the components of $\Sigma_{0}({\bf k},\omega)$ and
$\Sigma_{3}({\bf k}, \omega)$ have been treated within the framework
of the T-matrix approximation. However, it has been demonstrated
that the superconducting transition temperature is considerably
affected by the out-of-plane impurity scattering in spite of a
relatively weak increase of the residual resistivity \cite{eisaki}.
This reflects the fact that the superconducting pairing is very
sensitive to the out-of-plane impurity scattering, and then the
effect of the out-of-plane impurity scattering is always accompanied
by breaking of the superconducting pairs. In this case, the
out-of-plane impurities can be described as the elastic {\it
off-diagonal scatterers} or {\it pairing impurity scatterers}. In
particular, the modulation of the out-of-plane impurity scattering
potential observed in scanning tunneling microscopy experiments
\cite{pan} has a characteristic wavelength of a few lattice
spacings, this may arise because the impurities give rise to an
atomic-scale modulation of the charge carrier {\it pairing
potential} which causes larger, coherence length size fluctuations
in the out-of-plane impurity scattering potential \cite{nunner}.
Furthermore, the crude effect of the order parameter modulations on
the quasiparticle scattering by allowing the order parameter to be
modulated on the four bonds around the impurity has been estimated
\cite{graser} by adding the off-diagonal scattering potential,
\begin{eqnarray}
\hat{V}&=&\sum_{{\bf k},{\bf k}'}[V({\bf k})+V({\bf k}')]\tau_{1}
\nonumber\\
&=&{1\over 2}V_{0}\sum_{{\bf k},{\bf k}'}[({\rm cos}k_{x}-{\rm cos}
k_{y})+({\rm cos} k'_{x}-{\rm cos}k'_{y})]\tau_{1},~~~
\end{eqnarray}
to the phenomenological d-wave BCS Hamiltonian, then it was shown
that the scattering rate is largest at the antinode.

The exact form of the out-of-plane impurity scattering potential is
very important for a better understanding of the electronic
structure of the out-of-plane impurity-controlled cuprate
superconductors. In the following discussions, we determine the form
of the out-of-plane impurity scattering potential in terms of the
calculation of Dyson's equation. The potential which scatters the
electron is taken as summation of impurity potentials
$\tilde{V}=\sum_{l}V({\bf r}_{i} -{\bf R}_{l})$, where the summation
is over all impurity sites $l$, and then its Fourier transform is
obtained \cite{kohn,mahan} as $\tilde{V}({\bf q})=\rho_{i}V({\bf q})
\rho({\bf q})$, where
\begin{eqnarray}
\rho({\bf q})&=&\sum_{{\bf k}}h^{\dagger}_{{\bf k}+{\bf q}}h_{{\bf
k}},\\
\rho_{i}({\bf q})&=&\sum_{l}e^{i{\bf q}\cdot{\bf R}_{l}},
\end{eqnarray}
are the charge carrier density in the Nambu representation and the
impurity density, respectively. In the calculation of the
self-energy function induced by the impurity scattering, usually it
is assumed that the impurities are randomly located and that there
is no correlation between their positions \cite{kohn,mahan}. In this
case, the self-energy function can be obtained as
$\tilde{\Sigma}({\bf k}, \omega)= \tilde{\Sigma}^{(1)} ({\bf k},
\omega)+\tilde{\Sigma}^{(2)} ({\bf k},\omega)$ within the Born
approximation, with the corresponding first-order and second-order
self-energy functions are evaluated as \cite{kohn,mahan},
\begin{subequations}
\begin{eqnarray}
\tilde{\Sigma}^{(1)}({\bf k},\omega)&=&\rho_{i}\sum_{\bf k'}
\delta_{{\bf k'}=0}V({\bf k'})=\rho_{i}V(0),\\
\tilde{\Sigma}^{(2)}({\bf k},\omega)&=&\rho_{i}\sum_{{\bf k'},{\bf
k''}}\delta_{{\bf k'}+{\bf k''}=0}V({\bf k'})
\tilde{g}_{I}({\bf k}+{\bf k'},\omega)V({\bf k''})\nonumber\\
&=&\rho_{i}\sum_{{\bf k'}}V({\bf k'})\tilde{g}_{I}({\bf k}+{\bf k'},
\omega)V(-{\bf k'}),
\end{eqnarray}
\end{subequations}
where $\rho_{i}$ is the impurity concentration. As we have mentioned
above, the out-of-plane impurities are the off-diagonal scatterers.
Although their scattering has a very weak effect on the residual
resistivity for cuprate superconductors, a heavy effect on the
d-wave SC state is observed experimentally \cite{eisaki}. With these
considerations, we introduce the following out-of-plane impurity
scattering potential,
\begin{eqnarray}
\tilde{V}=\sum_{{\bf k'}}V({\bf k'})\tau_{1}=V_{0}\sum_{{\bf k'}}
[{\rm cos}k'_{x}-{\rm cos}k'_{y}]\tau_{1}.
\end{eqnarray}
In this case, $V(0)=V_{0}[{\rm cos}(0)-{\rm cos}(0)]=0$ (then
$\tilde{\Sigma}_{1}({\bf k},\omega)=0$), and $\tilde{\Sigma}({\bf k}
,\omega)=\tilde{\Sigma}^{(2)}({\bf k},\omega)$. This form of the
out-of-plane impurity scattering potential in Eq. (12) is very
similar to that in Eq. (8) used in Ref. \onlinecite{graser}, and the
scattering rate is also largest at the antinode. This is indeed
confirmed by the quantitative characteristics presented in the
following section. With the help of the impurity scattering
potential in Eq. (12), the components of the charge carrier
self-energy function $\tilde{\Sigma}({\bf k},\omega)$ are obtained
explicitly as,
\begin{subequations}
\begin{eqnarray}
\Sigma_{0}({\bf k},\omega)&=&\rho_{i}{1\over N}\sum_{{\bf k'}}
|V({\bf k'})|^{2}{g}_{I0}({\bf k'+k},\omega)\nonumber\\
&=&\rho_{i}{1\over N}\sum_{{\bf k'}}|V({\bf k'-k})|^{2}
{g}_{I0}({\bf k'},\omega) ,
\end{eqnarray}
\begin{eqnarray}
\Sigma_{3}({\bf k},\omega)&=&-\rho_{i}{1\over N}\sum_{{\bf k'}}|
V({\bf k'})|^{2}{g}_{I3}({\bf k'+k},\omega)\nonumber\\
&=&-\rho_{i}{1\over N}\sum_{{\bf k'+k}}|V({\bf k'-k})|^{2}
{g}_{I3}({\bf k'},\omega),~~~~\\
\Sigma_{1}({\bf k},\omega)&=&\rho_{i}{1\over N}\sum_{{\bf k'}}|
V({\bf k'})|^{2}{g}_{I1}({\bf k'+k},\omega)\nonumber\\
&=&\rho_{i}{1\over N} \sum_{{\bf k'}}|V({\bf k'-k})|^{2}
{g}_{I1}({\bf k'},\omega) .
\end{eqnarray}
\end{subequations}

In the charge-spin separation fermion-spin theory \cite{feng2}, the
electron diagonal and off-diagonal Green's functions are the
convolutions of the spin Green's function \cite{guo,feng,feng1}
$D^{-1}({\bf p},\omega)=(\omega^{2}-\omega_{{\bf p}}^{2})/B_{{\bf p}
}$ and the charge carrier diagonal and off-diagonal Green's
functions in Eq. (7), respectively. These convolutions reflect the
charge-spin recombination \cite{anderson1}. Following the previous
discussions \cite{guo,feng,feng1}, we can obtain the electron
diagonal and off-diagonal Green's functions in the present case.
Then the electron spectral function from the electron diagonal
Green's function is found explicitly as,
\begin{widetext}
\begin{eqnarray}
A({\bf k},\omega)&=&{1\over N}\sum_{\bf p}{B_{\bf p}\over
2\omega_{\bf p}}\{[n_{B}(\omega_{\bf p})+n_F(\omega_{\bf p}-
\omega)]A_{h}({\bf p}-{\bf k},\omega_{\bf p}-\omega)\nonumber\\
&-&[n_{B}(-\omega_{\bf p})+n_{F}(-\omega_{\bf p}-\omega)]A_{h}({\bf
p}-{\bf k},-\omega_{\bf p}-\omega)\},~~~~~
\end{eqnarray}
\end{widetext}
where $A_{h}$ is the charge carrier spectral function, which can be
expressed as $A_{h}=-2{\rm Im}g^{dia}_{I}({\bf k},\omega)$, with
$g^{dia}_{I}$ obtained from Eq. (7) as,
\begin{widetext}
\begin{eqnarray}
g^{dia}_{I}({\bf k},\omega)=Z_{hF}{\omega-\Sigma_{0}({\bf k},
\omega)+\bar{\xi}_{{\bf k}}+\Sigma_{3}({\bf k},\omega)\over
[\omega-\Sigma_{0}({\bf k},\omega)]^{2} -[\bar{\xi}_{{\bf k}}+
\Sigma_{3}({\bf k},\omega)]^2-[\bar{\Delta}_{hZ}({\bf k})+
\Sigma_{1}({\bf k},\omega)]^2}.
\end{eqnarray}
\end{widetext}

\section{Electronic structure of the out-of-plane impurity-controlled
cuprate superconductors}

Experimentally, it has been shown that the average of the
next-nearest neighbor hopping $t'$ is not appreciably affected by
the out-of-plane impurities \cite{hashimoto}. In this case, the
commonly used parameters in this paper are chosen as $t/J=2.5$ and
$t'/t=0.3$. We are now ready to discuss the electronic structure of
the out-of-plane impurity-controlled cuprate superconductors and the
related superconducting gap. In cuprate superconductors, the
information revealed by ARPES experiments \cite{shen} has shown that
around the nodal [$\pi/2,\pi/2$] and antinodal [$\pi,0$] points of
the Brillouin zone contain the essentials of the whole low energy
quasiparticle excitation spectrum. In this case, we have performed a
calculation for the electron spectral function $A({\bf k},\omega)$
in Eq. (14) at both nodal and antinodal points. The results at (a)
the nodal point and (b) the antinodal point with the impurity
concentration $\rho_{i}=0.001$ (solid line), $\rho_{i}=0.002$
(dashed line), and $\rho_{i}=0.003$ (dotted line) under the impurity
scattering potential with $V_0=50J$ for the charge carrier doping
concentration $\delta=0.15$ are plotted in Fig. 1. For comparison,
the corresponding ARPES experimental results \cite{hashimoto} for
the out-of-plane impurity-controlled cuprate superconductors
$Ln$-Bi2201 in the superconducting state are also presented in Fig.
1 (inset). Our results show that the quasiparticle peak is strongly
dependent on the impurity concentration, and the peaks at both nodal
and antinodal points are suppressed due to the presence of impurity
scattering. At the nodal point, there is a sharp superconducting
quasiparticle peak near the Fermi energy, however, although the peak
at the high impurity concentration is dramatically reduced compared
to that at the low impurity concentration, the position of the
leading-edge mid-point of the electron spectral function remains
almost unchanged. In particular, the position of the leading-edge
mid-point of the electron spectral function reaches the Fermi level,
indicating that there is no superconducting gap. On the other hand,
the spectral intensity from the Fermi energy down to $\sim -1.1J$
decreases as the impurity concentration increases at the antinodal
point, this is the same case as that at the nodal point. However,
the position of the leading-edge mid-point of the electron spectral
function is shifted towards higher binding energies with increasing
impurity concentration, this is in contrast with the behavior
observed at the nodal point, and indicates the presence of the
superconducting gap. The present results also show that the effect
of the out-of-plane impurity scattering is stronger at the antinodal
point than at the nodal one, in qualitative agreement with the
experimental results \cite{kondo,hashimoto}.

\begin{figure}[t]
\begin{center}
\leavevmode
\includegraphics[clip=true,width=1.0\columnwidth]{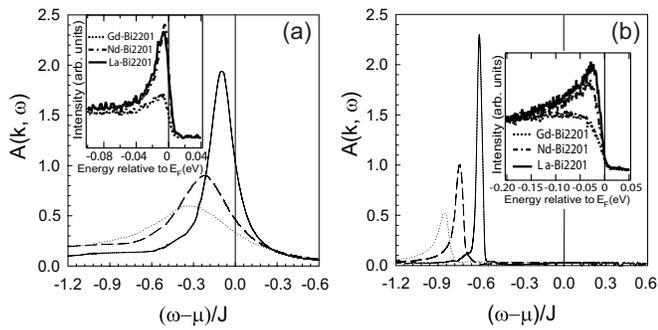}
\caption{The electron spectral function at (a) the nodal point and
(b) the antinodal point with $\rho_{i}=0.001$ (solid line),
$\rho_{i}=0.002$ (dashed line), and $\rho_{i}=0.003$ (dotted line)
for $V_{0}=50J$ at $\delta=0.15$. Inset: the corresponding
experimental results taken from Ref. \onlinecite{hashimoto}.}
\end{center}
\end{figure}

\begin{figure}[t]
\begin{center}
\leavevmode
\includegraphics[clip=true,width=0.9\columnwidth]{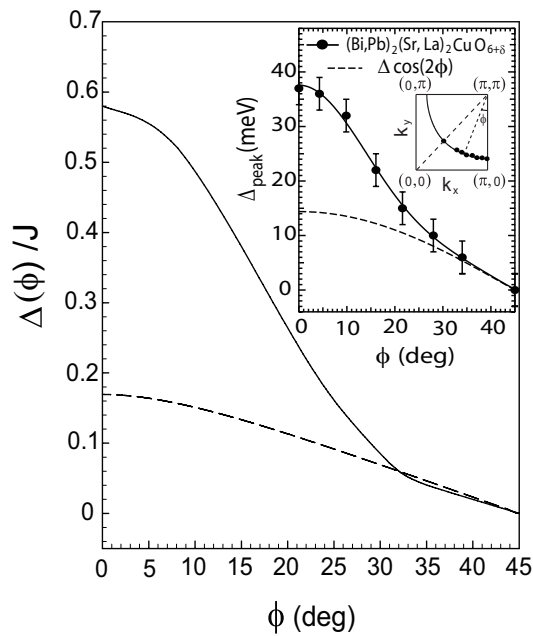}
\caption{The superconducting gap as a function of the Fermi surface
angle $\phi$ with $\rho_{i}=0$ (dashed line) and $\rho_{i}=0.001$
(solid line) for $V_{0}=50J$ at $\delta=0.15$. Inset: the
corresponding experimental results taken from Ref.
\onlinecite{kondo}.}
\end{center}
\end{figure}

The behavior of the electron spectrum in Fig. 1 indicates an
enhancement of the superconducting gap in the antinodal region by
the impurity scattering. To show this point clearly, we have
calculated the electron spectral function $A({\bf k},\omega)$ along
the direction $[\pi,0]\rightarrow [\pi/2,\pi/2]$, and then employed
the shift of the leading-edge mid-point as a measure of the
magnitude of the superconducting gap at each momentum just as it has
been done in the experiments \cite{kondo,hashimoto}. The results for
the extracted superconducting gap as a function of the Fermi surface
angle $\phi$, defined in the inset, with the impurity concentration
$\rho_{i}=0$ (dashed line) and $\rho_{i}=0.001$ (solid line) under
the impurity scattering potential with $V_0=50J$ for the charge
carrier doping concentration $\delta=0.15$ are plotted in Fig. 2 in
comparison with the corresponding ARPES experimental results for the
out-of-plane impurity-controlled cuprate superconductor
(Bi,Pb)$_{2}$(Sr,La)$_{2}$CuO$_{6+\delta}$ in the superconducting
state \cite{kondo} (inset). It is clearly shown that the
superconducting gap $\Delta$ increases with the Fermi surface angle
decreasing from 45$^{{\rm o}}$ (node) to 0$^{{\rm o}}$ (antinode).
Although the superconducting gap in the presence of the impurity
scattering is basically consistent with the d-wave symmetry, it is
obvious that there is a strong deviation from the monotonic d-wave
form around the antinodal region. In particular, this strong
deviation is mainly caused by a remarkable enhancement of the
superconducting gap value around the antinodal region, in
qualitative agreement with the experimental results
\cite{kondo,hashimoto}. In other words, the superconducting gap
around the antinodal region is strongly enhanced by the impurity
scattering, whereas around the nodal region its value remains the
same. As a consequence, the well pronounced deviation from the
monotonic d-wave superconducting gap form in the out-of-plane
impurity-controlled cuprate superconductors is mainly caused by the
effect of the out-of-plane impurity scattering. This is also the
reason why the superconducting gap function for very high quality
samples of the cuprate superconductor La$_{1-x}$Sr$_{x}$CuO$_{4}$
has a monotonic d-wave form \cite{shi1}.

\begin{figure}[t]
\begin{center}
\leavevmode
\includegraphics[clip=true,width=0.9\columnwidth]{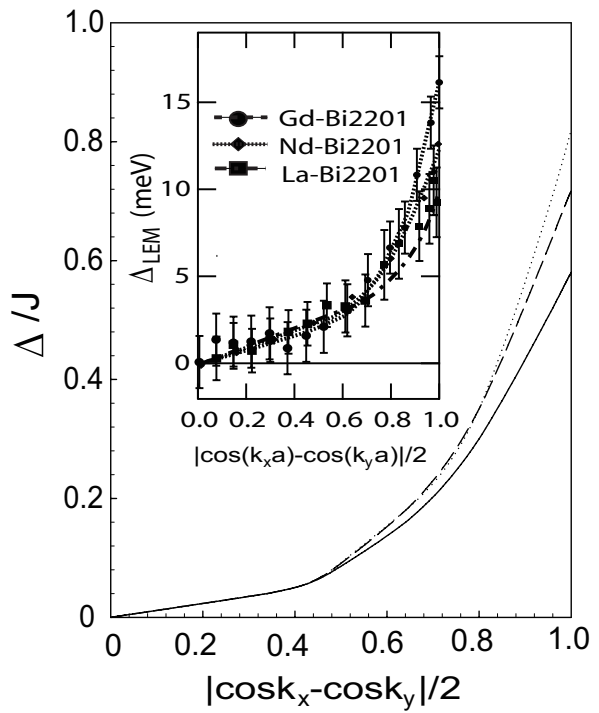}
\caption{The superconducting gap as a function of $[{\rm
cos}k_{x}-{\rm cos}k_{y}]/2$ with $\rho_{i}=0001$ (solid line),
$\rho_{i}=0.002$ (dashed line), and $\rho_{i}=0.003$ (dotted line)
for $V_0=50J$ at $\delta=0.15$. Inset: the corresponding
experimental results taken from Ref. \onlinecite{hashimoto}.}
\end{center}
\end{figure}

For a better understanding of the impurity concentration dependence
of the deviation from the monotonic d-wave superconducting gap
function, we have made a series of calculations for the
superconducting gap at different impurity concentration levels, and
the results of the superconducting gap as a function of the
monotonic d-wave function $[{\rm cos}k_{x}-{\rm cos}k_{y}]/2$ with
the impurity concentration $\rho_{i}=0001$ (solid line),
$\rho_{i}=0.002$ (dashed line), and $\rho_{i}=0.003$ (dotted line)
under the impurity scattering potential with $V_0=50J$ for the
charge carrier doping concentration $\delta=0.15$ are plotted in
Fig. 3 in comparison with the corresponding ARPES experimental
results for the out-of-plane impurity-controlled cuprate
superconductors Ln-Bi2201 \cite{hashimoto} (inset). Obviously, our
results show that the magnitude of the deviation from the monotonic
d-wave superconducting gap form around the antinodal region
increases with increasing impurity concentration, in qualitative
agreement with the experimental results \cite{kondo,hashimoto}. This
strong out-of-plane impurity effect in the antinodal region is also
consistent with scanning tunneling spectroscopy results
\cite{sugimoto}, where the average of the superconducting gap size,
which corresponds to the antinodal superconducting gap in the ARPES
spectra, increases with increasing impurity concentration.

Within the framework of the kinetic energy driven superconducting
mechanism \cite{feng1} in the presence of the out-of-plane
impurities, our present results show that the out-of-plane impurity
scattering potential (12) in which the impurities modulate the pair
interaction locally give qualitative agreement with respect to the
main features observed in the ARPES measurements on the out-of-plane
impurity-controlled cuprate superconductors in the superconducting
state. Although this out-of-plane impurity effect in cuprate
superconductors can also be discussed starting directly from a
phenomenological d-wave BCS formalism \cite{graser,nunner}, in this
paper we are primarily interested in exploring the general notion of
the effects of the out-of-plane impurity scatterers in the kinetic
energy driven cuprate superconductors in the superconducting state.
The qualitative agreement between the present theoretical results
and ARPES experimental data also indicates that the presence of the
out-of-plane impurities has a crucial impact on the electronic
structure of cuprate superconductors. On the other hand, we
emphasize that the quasiparticle scattering rate in the antinodal
region is strongly increased by the impurity scattering potential
(12), while the nodal quasiparticles are very weakly scattered by
the impurity scattering potential (12), this is why the
superconducting transition temperature is considerably affected by
the out-of-plane impurity scattering in spite of a relatively weak
increase of the residual resistivity \cite{eisaki}, since the
transport properties are mainly governed by the quasiparticles in
the nodal region.

\section{Summary}

In conclusion, we have shown very clearly in this paper that if the
out-of-plane impurity scattering is taken into account within the
framework of the kinetic energy driven d-wave superconductivity
\cite{feng1}, the quasiparticle spectrum of the $t$-$J$ model
calculated based on the off-diagonal impurity scattering potential
(12) per se can correctly reproduce some main features found in the
ARPES measurements on the out-of-plane impurity-controlled cuprate
superconductors in the superconducting state \cite{kondo,hashimoto}.
In the presence of the out-of-plane impurities, although both sharp
superconducting coherence peaks around the nodal and antinodal
regions are suppressed, the effect of the impurity scattering is
stronger in the antinodal region than that in the nodal region, this
leads to a strong deviation from the monotonic d-wave
superconducting gap form in the out-of-plane impurity-controlled
cuprate superconductors.

Finally, we have noted that within a phenomenological BCS approach,
the electron spectral properties of the underdoped cuprates as
resulting from a momentum dependent pseudogap in the {\it normal
state} have been discussed \cite{sangiovanni}, where a normal state
pseudogap function deviating from the monotonic d-wave pseudogap
form has been used to fit the ARPES experimental data in the normal
state. It has been shown \cite{kastner,shen95} that there are some
subtle differences for different families of underdoped cuprates in
the normal state, and therefore it is possible that the pseudogap in
the normal state is effected by the impurity scattering as well.

\acknowledgments

This work was supported by the National Natural Science Foundation
of China under Grant No. 10774015, and the funds from the Ministry
of Science and Technology of China under Grant Nos. 2006CB601002 and
2006CB921300.



\end{document}